# Oxygen-vacancy Mediated Deterministic Domain Distribution at the Onset of Ferroelectricity

Asaf Hershkovitz,[†,1,2] Elangovan Hemaprabha,[†,1,2] Doaa Khorshid,[1,2] Liyang Ma,[3] Shi Liu,[3,4] Shai Cohen,[5,7] Yachin Ivry[1,2,6 *]

[1] Department of Materials Science and Engineering, Technion – Israel Institute of Technology, Haifa 3200003, Israel.

[2] Solid State Institute, Technion – Israel Institute of Technology, Haifa 3200003, Israel.

[3] Key Laboratory for Quantum Materials of Zhejiang Province, School of Science, Westlake University, Hangzhou 310024, Zhejiang Province, China.

[4] Institute of Natural Sciences, Westlake Institute for Advanced Study, Hangzhou 310024, Zhejiang Province, China.

[5] Nuclear Research Center – Negev, Beer Sheva 84190, Israel.

[6] Department of Materials Science and Engineering, University of California, Berkley, CA 94720, USA.

[7] Nuclear Engineering Department, University of California, 4153 Etcheverry Hall Berkeley, Berkeley, CA 94720, USA.

[†] These authors contributed equally to the work.

[*]Correspondence to: ivry@technion.ac.il, +972-4-8294561

## Abstract

Ferroelectric domains are mesoscale structures that mediate between synchronized atomic-scale ion displacements and switchable macroscopic polarization. Here, we evaluated the randomness of the domain distribution at the onset of ferroelectricity. First-principle calculations combined with atomic-scale imaging demonstrate that oxygen vacancies that serve as pinning sites for the ferroic domain walls remain immobile above the Curie temperature. Thus, upon cooling to a ferroelectric state, these oxygen vacancies dictate reproducible domain-wall patterning. Domain-scale imaging with variable-temperature piezoresponse force microscopy confirmed the memory effect, questioning the spontaneity of domain distribution under thermotropic transitions.

## 1. Introduction

When cooling a proper ferroelectric material down, below the Curie temperature ($T_{\mathrm{C}}$), the crystal symmetry is reduced to a non-centrosymmetric structure and polarization domains evolve [1,2]. A seminal example is the cubic-to-tetragonal transition in barium titanate [3,4]. Following Neumann's principle [5], the three orthogonal crystallographic orientations of the tetragonal structure dictate six possible polarization domain orientations that are parallel to the faces. Presumably, upon repeating cooling-heating cycles around the ferroelectric transition, the polarization orientation at each area in the material should pick up an arbitrary orientation within this six-possibility space. Likewise, the position of the domain walls should also be random, at least at the macroscopic scale. That is, there should be no memory effect and no significant correlation between the domain distributions during repeating transitions.

There are two main factors that can introduce a certain degree of memory during the phase transformation, deteriorating the randomness of the polarization distribution. First, the polarization is restricted to the long axis of the tetragonal structure, giving rise to two possible orientations along this axis. However, the long tetragonal axis itself can orientate in three different directions: one out-of-plane and two in-plane. Similar to the common alignment of dipole moments in neighboring unit cells that form the polarization domains, neighboring unit cells align their crystallographic axis in a common direction, giving rise to ferroelastic domains. Thus, the polarization domain distribution is limited by the ferroelastic domains. Following Roytburd's scaling law [6], ferroelastic domains have a characteristic periodicity ($w$) that depends on the strain conditions ($S$) and material thickness ($d$):

$$S = w^2 d \tag{1}$$

Yet, Roytburd's model does not impose any restrictions on the position and orientation of the domain walls. Thus, examining the ferroelastic domain redistribution around $T_{\mathrm{C}}$ can help evaluate the randomness in mesoscale-polarization distribution in ferroelectrics.

A second reason for assuming a possible lack of randomness in the polarization distribution stems from conceptual equivalence between domain distribution due to thermal and electric-field excitations. That is, domains redistribute not only around $T_{\mathrm{C}}$ but also under an external electric field that is larger than the coercive value. Recent studies indicate on a certain equivalence between thermal-induced and electric-field-induced domain redistribution, such as the size of domain nuclei [7,8] and the interplay between long- and short- range interactions [9–12]. While the information regarding domain redistribution around the phase transition is limited, much effort has been put in understanding retention and aging effects in ferroelectric domains under electric excitations [13–16], mainly due to the significance of these materials to the data-storage industry. Under repeating polarization switching with an electric field, the domains eventually become pinned and are not free to move as they were in the beginning [10]. The common concept is that the domains are pinned by local immobile point defects [17], with a recent focus on the possibility that these defects are oxygen vacancies [18,19]. Thus, it is interesting to examine whether such domain-pinning effects take place also at thermal-induced domain redistribution. Moreover, due to the difficulties involved in observing point defects and domain walls directly, either individually or simultaneously, there is only limited experimental evidence to support the related hypotheses [20,21].

A careful look into several very recent works reveals that indeed, domains have a certain memory also upon thermal switching [22–24]. Additional focus has been given to attempts to utilize the large strain involved in ferroelastic domain switching in shape-memory effects under thermal-stress excitations [25]. This effect does not require necessarily identical domain pattern upon repeating phase-transformation cycles, but the repeating global strain indicates on an averaged similarity in the macroscopic scale domain structure [25]. However, the basic understanding of whether the polarization distribution in ferroelectrics is arbitrary or not has yet remained unclear. Hence, a methodical examination of randomness in domain distribution and the elusive factors that affect it in the realm of the ferroelectric phase transformation is needed.

Here, we combined theoretical and multiscale experimental analysis to show that under sequential thermal-induced transitions, the domain distribution is not random. Rather, there is a significant correlation between the re-formed ferroelectric domain structures. Using variable-temperature atomic-force microscopy (AFM) and piezoresponse force microscopy (PFM) around $T_C$, it is shown that the correlation between a set of domain structures taken from eight repetitive transition cycles is as high as 0.81±0.05 in comparison to zero for the case of simulated hypothetically arbitrary domain formation, even under the restrictions of Roytburd's scaling law. Domain repeatability around the ferroelectric-ferroelectric tetragonal-orthorhombic transition is also demonstrated. Ab-initio calculations proposed that oxygen vacancies that are formed during the domain-wall formation remain immobile even upon heating the ferroelectric above $T_C$, hence serving as pinning sites to the reproducible domain pattern. Recent development in the ability to image oxygen vacancies in $BaTiO_3$ by means of advanced electron microscopy [31] was used to validate the ab-initio results, revealing that indeed, oxygen vacancies are formed at the domain walls and they both remain in place even upon heating the ferroelectric above $T_C$ and cooling it down back to room temperature.

## 2. Results and Discussion

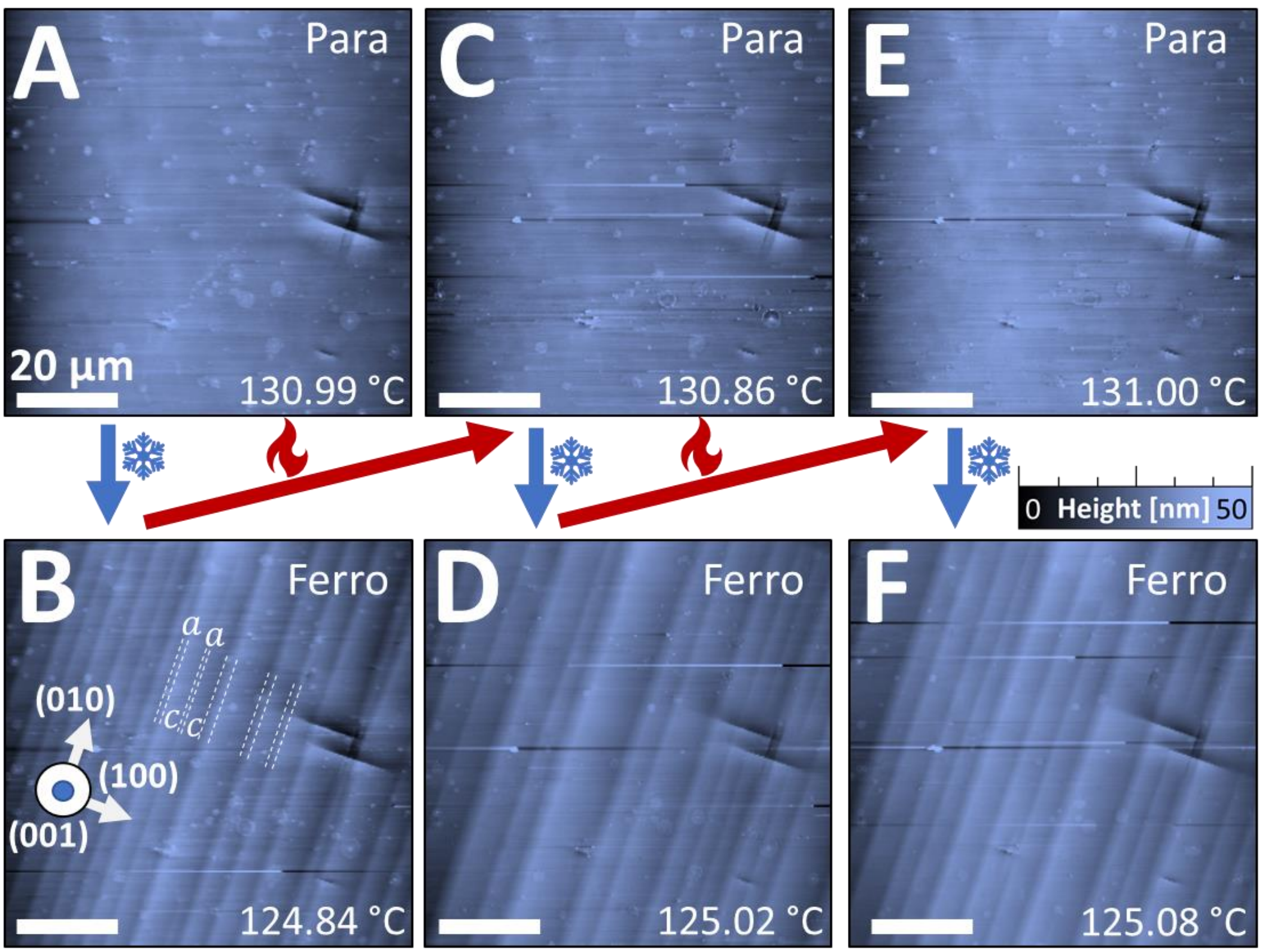

**Figure 1| Domain repeatability around the paraelectric-ferroelectric phase transition in single-crystal $BaTiO_3$.** (**A**) An atomic-force micrograph of a large area (80 × 80 µm$^2$) obtained above $T_C$, showing no domains. (**B**) Periodically alternating *a*-*c* domains were formed upon cooling the crystal below $T_C$. (**C**) Heating the ferroelectric again above $T_C$ was accompanied by the disappearance of the domains that (**D**) redistributed very similarly to their previous arrangement (B) upon cooling the sample back to the ferroelectric state. Repeating the (**E**) heating – (**F**) cooling cycle yielded similar results, which was reproducible also in an entire set of eight sequencing phase-transformation cycles (see Figure S1 for details). The scan temperature is specified in each image.

First, the mesoscale behavior was examined. That is, the repeatability of large-scale polarization domain distribution was evaluated. Here, AFM [26] and PFM [27] were used to image the ferroelectric and ferroelastic domain distribution in bulk single-crystal $BaTiO_3$. A recently developed *in-situ* temperature control during PFM imaging [9,28] was used to drive the material thermally through periodic phase transitions and image the resultant domain distribution. Figure 1 shows representative domain distribution during repeating eight phase-transition cycles. Figure 1A shows that above $T_C$, at the paraelectric cubic phase, no domains were observed, as expected. Upon cooling the material below $T_C$, a periodic alternating striped 90° domain structure was formed (Figure 1B). While the periodicity followed roughly Roytburd's scaling law (Equation 1) [6,29], the domain-wall position and orientation were presumably random. Heating the crystal again above $T_C$ (Figure 1C, no domains were observed) and cooling it down back below $T_C$ (Figure 1D) resulted in a domain distribution that was very similar to the original structure (Figure 1B) by means of domain-wall orientation, position and periodicity. A phase-transition cycle (Figures 1E-F) yielded a similar result, and so did the entire eight-cycle set (Figure S1).

Although the domain repeatability is clearly observed in Figure 1 even with a naked eye, we wanted to evaluate the effect quantitatively. Hence, the correlation among the eight different sequencing domain distributions that are presented in Figure 1 was extracted. Here, the height profile of a cross section that is perpendicular to the domain walls from each image was used to compare the domain distributions. The different cross sections were taken from exactly the same area (*i.e.*, the cross sections overlapped). We then extracted the correlation value between any two cross sections of the eight images, so that 28 different correlation values were calculated. Choosing this method allowed us to ignore the inner order of the images and hence to examine the randomness of the domain distribution in the entire data set. Figure 2A shows the correlation distribution of the 28 comparisons. Fitting the data to a normal-distribution function resulted in a correlation value that ranges between 0.72-0.90 with a mean value of 0.81±0.01 and 0.05 standard deviation of the entire distribution. Hence, the correlation value was determined as $r = 0.81 \pm 0.05$. This result indicates that the overlap in domain distribution of two random experiments in the set ($r^2$) is 64%. The significance of this result can be extracted from the statistical *p*-test, which was found to be $< 0.0005$. To further examine the significance of this value in the context of the physical phenomenon and not only in the formal statistical test, we formed a simulated dataset of 10,000 random domain distributions that correspond to a hypothetical 10,000 repeating experiments. The correlation of this 'arbitrary' domain-distribution dataset was then calculated. Here, we assumed that the domain width is within the range of 2 - 5 μm, which agrees with Equation 1 as well as with the experimental observations (Figure 1). However, the exact domain widths and the exact position of the domain walls were set randomly for each of the 10,000 experiments. Likewise, the domain shape and height pick were set according to the experimental observations (Figure 2B). Then, each of the domain distributions within the simulated dataset were compared to the experimental observation that resulted in the highest averaged correlation within the eight experimental images (the domain distribution from Figure S1A). Figure 2C shows that the simulation resulted in a mean correlation of $r = 0.0 \pm 0.1$ (within the error of the standard deviation of the distribution). Thus, the above conservative value $r = 0.81 \pm 0.05$ demonstrates clear significance of highly correlated domain distributions that are not random.

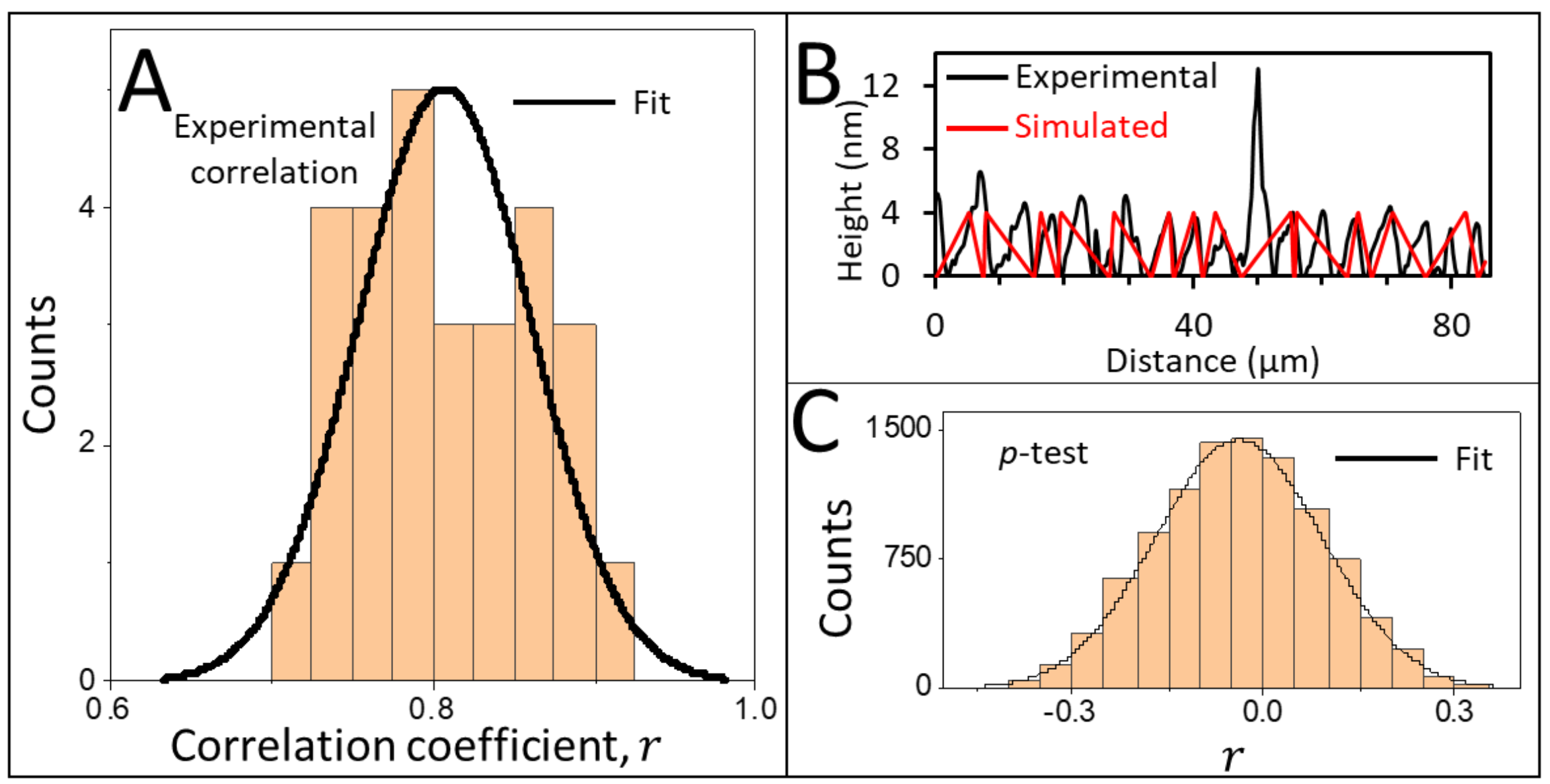

**Figure 2| Correlation of ferroelectric domain redistribution under repeating paraelectric-ferroelectric transitions.** (**A**) A histogram of correlation coefficient values for 28 domain-pattern comparisons of eight consecutive transitions. The normal-distribution fit peaks at 0.81±0.05, showing high correlation and low randomness. (**B**) Line-profile along the highest correlated domain pattern (see Figure S1B) and a representative simulated signal (among 10,000 generated signals). (**C**) Correlating the experimental profile in (B) with 10,000 simulated signals resulted in $r = 0.0$, illustrating the significance of the correlation values in (A).

To complete the experimental evaluation of randomness in domain distribution, we examined another ferroelectric phase transformation that allowed us look at the domain distribution both above and below the transition. Here, the ferroelectric-ferroelectric transition that is accompanied by a symmetry lowering from tetragonal to orthorhombic crystal structure was chosen. Figure 3 shows the similarity of the tetragonal domain redistribution as well as of the orthorhombic domain distribution during the thermal-driven phase-transition cycles as was revealed with PFM.

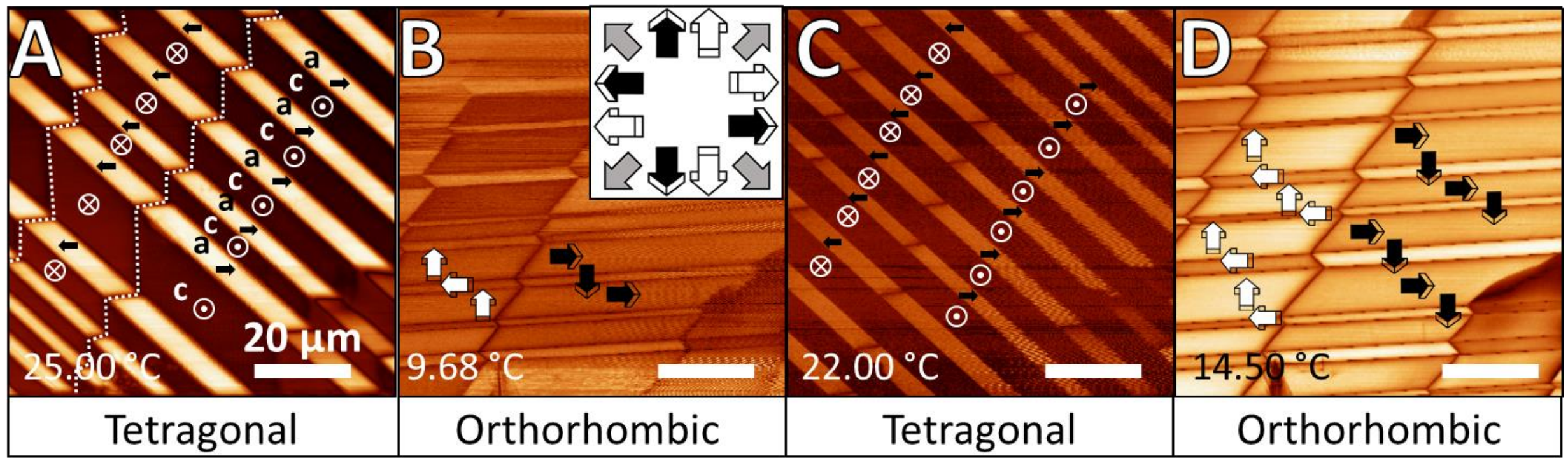

**Figure 3| Domain reproducibility around a ferroelectric-ferroelectric phase transition in single-crystal $BaTiO_3$**. (**A**) PFM micrograph of the domain distribution in the tetragonal phase and (**B**) at the same area after cooling the crystal to the orthorhombic phase. Sequential (**C**) heating and (**D**) cooling the material again to the tetragonal and orthorhombic phases, respectively, gave rise to a repeated domain distribution. The scan temperature is specified in the images. The 3D phase and amplitude mode signals of this experiment as well as the simultaneously imaged topography that are complementary to the lateral amplitude images that displayed here are given in Figure S2.

Next, the atomic-scale origin of the domain-wall pinning during repeating phase-transition cycles was examined. Recent advances in electron microscopy proved useful for revealing atomic-scale static and dynamic behavior of point defects in general [30] and individual oxygen vacancies at ferroelectric domain walls in particular [31]. These achievements are possible when the transmission-electron microscopy (TEM) imaging methods of high-angle annular dark-field (HAADF), differential phase contrast (DPC) and integrated DPC (iDPC) imaging are performed simultaneously. Here, we used this method to examine pinning effects under heating-cooling cycles. First, domains were formed by means of *in-situ* contactless electric-field application in a single-crystal $BaTiO_3$, following a previously introduced experimental protocol [31–33]. Figure 4A shows a set of four oxygen vacancies along a 90° domain wall and another vacancy at an adjacent domain wall. The crystal was then heated (*in-situ*) to 180 °C, above the 140 °C Curie temperature in these materials [7] and lastly was cooled down back to room temperature. Figure 4B shows that in the new state, the domain walls were formed at the same location that they were observed prior to the phase transition. Likewise, the oxygen vacancies remained in their place in comparison to Figure 4A. Complementary TEM images and related analyses are given in Figures S3 and S4.

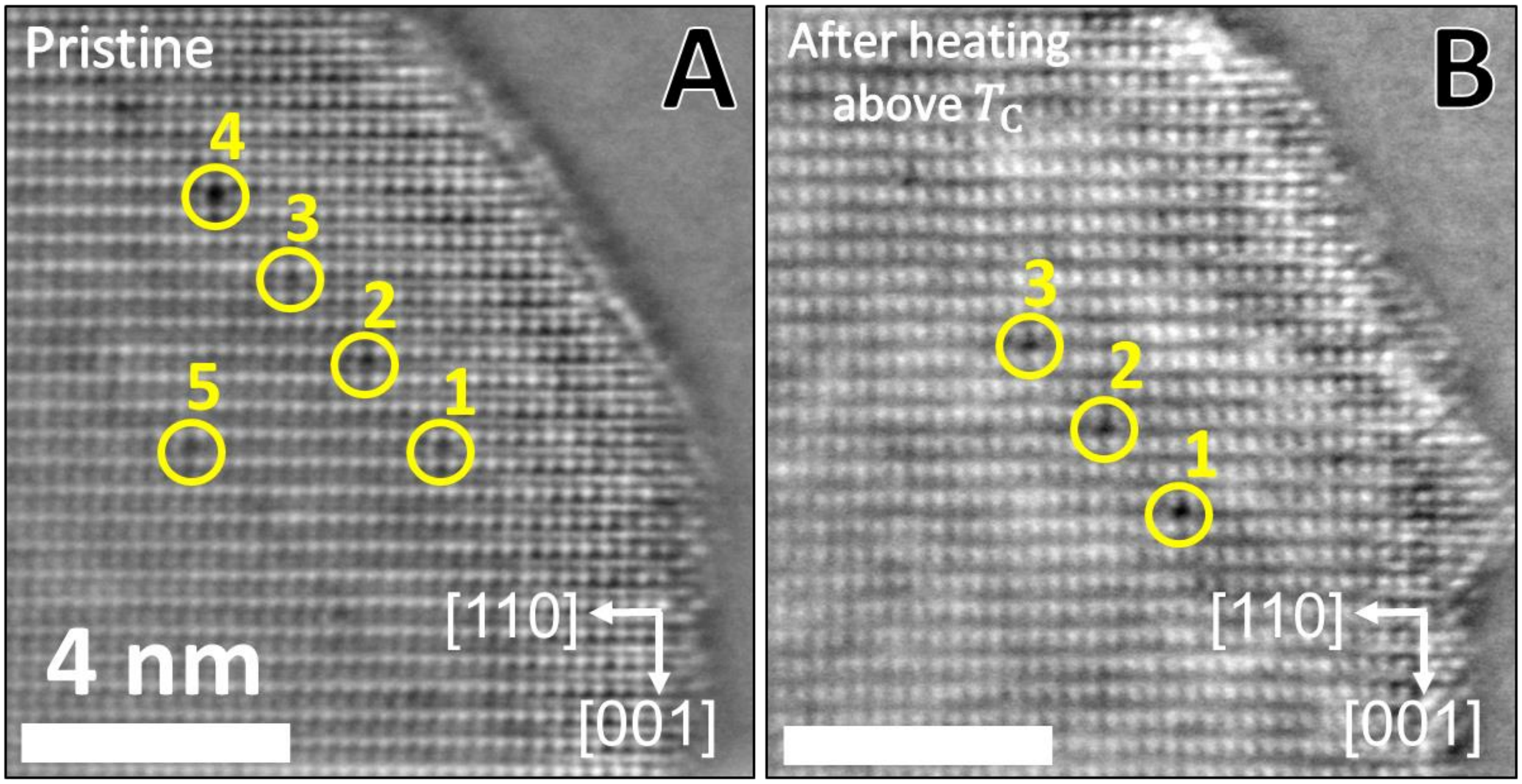

**Figure 4| Pinned point defects and domain walls during a ferroelectric phase-transition cycle.** (**A**) 90° domain walls (parallel lines with a darker contrast along the $[\bar{1}11]$ direction, see also Figure S3) and a series of oxygen vacancies (designated and numbered) in the domain walls (see Ref 31 for the detailed experimental protocol), revealed with atomic-scale iDPC imaging of a $BaTiO_3$ single crystal. (**B**) The same area after heating the crystal above $T_C$ and cooling it down back to room temperature demonstrates that the domain-wall location has remained unchanged. Moreover, at least three oxygen vacancies remained in the same position as in (A) at the domain wall. Complementary TEM images and related analyses are given in Figures S3 and S4. Large-scale iDPC and DPC micrographs of (A) and (B) as well as the position of the domain walls are given in Figure S3.

To support the experimental results of oxygen-vacancy-induced pinned domain walls under phase-transition cycles, the oxygen-vacancy mobility in the ferroelectric and paraelectric phases was examined. Here, the oxygen-vacancy diffusion in $BaTiO_3$ was modeled based on density functional theory (DFT) [34–36]. Figure 5 shows a 0.95 eV calculated diffusion barrier for an oxygen vacancy at the tetragonal phase, which agrees with previous DFT calculations [37–39]. This energy profile was calculated for the diffusion of oxygen vacancies between the axial and equatorial sites relative to the polarization direction and is the lowest value that was obtained from the various calculations (additional energy profiles for other diffusion paths are presented in Figure S5). Note that the computed activation energies agree well with previous experimental values obtained with different techniques, including vacancy electromigration (~1 eV), time-of-flight secondary ion mass spectrometry (~0.70 eV), and electron paramagnetic resonance (~0.9 eV) [40]. Interestingly, the oxygen diffusion coefficient ($D_{v_O}$) in $BaTiO_3$ was found to depend sensitively on the temperature, particularly around $T_C$. For example, Kessel *et al.* [40] estimated that $D_{v_O}$ changes from $10^{-21}$ $m^2s^{-1}$ at 400 K to $10^{-14}$ $m^2s^{-1}$ at 466 K, corresponding to angstrom-level and micrometer-level migration length over a one-second time scale. The DFT calculations (Figure 5) indicate that the barrier is maintained also at the cubic phase (albeit with a slightly higher value of 1.05 eV), suggesting that the transformation from tetragonal to cubic has only a minor impact on the activation energy of oxygen-vacancy diffusion. That is, already-existing oxygen vacancies can serve as nucleation sites to the newly formed domains, effectively pinning the domains during the heating-cooling cycle when the heating-cooling cycles are controlled accurately [41].

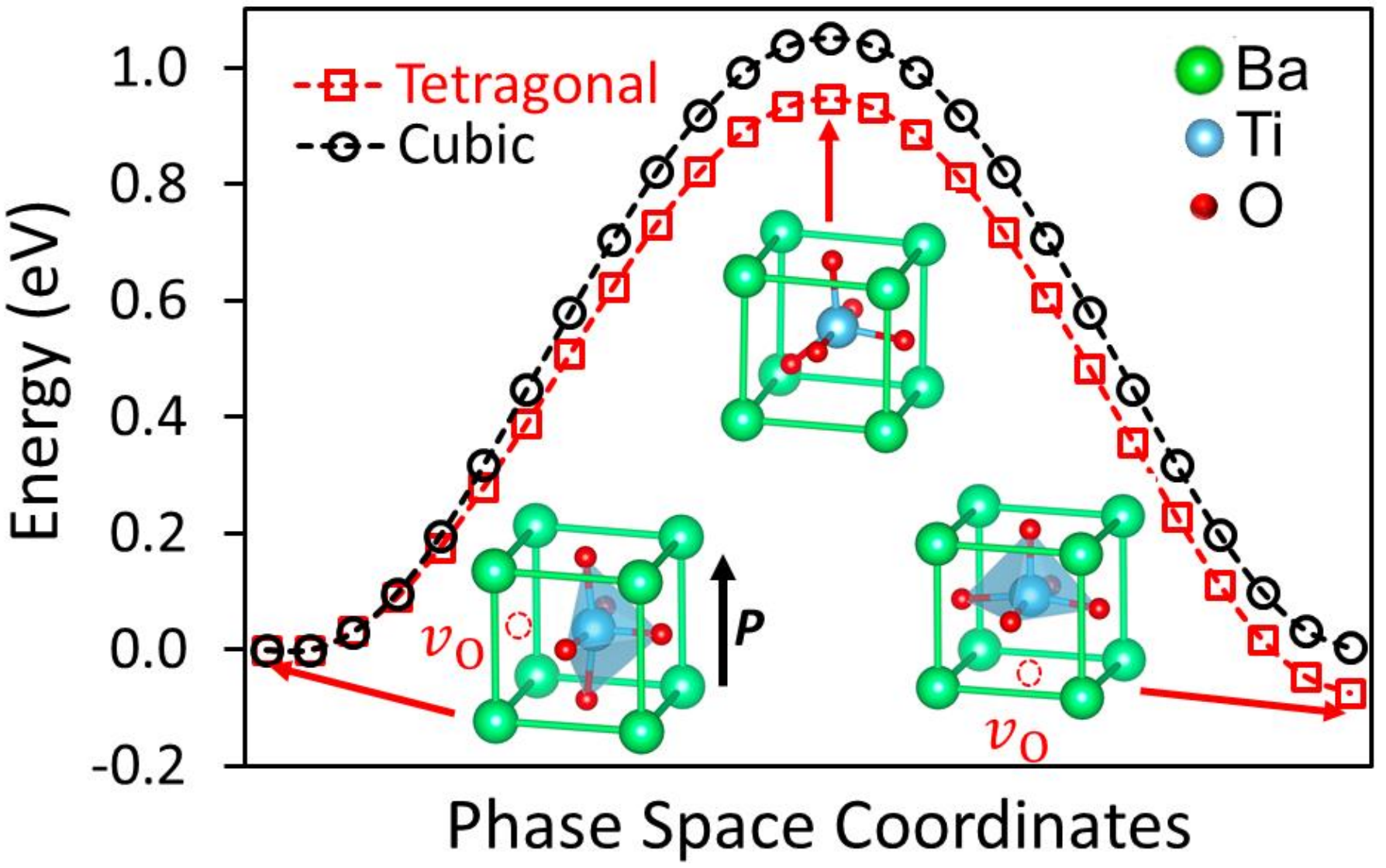

**Figure 5| Energy profiles for oxygen-vacancy diffusion in tetragonal and cubic $BaTiO_3$.** An oxygen vacancy diffuses between the equatorial site (left inset) and the axial site (right inset) relative to the polarization orientation. The diffusion barrier in the tetragonal phase is 0.95 eV, comparable with the 1.05 eV barrier in the cubic phase. The insets show the structures of the unit cell containing one oxygen vacancy in a 2×2×2 supercell. The supercell with axial oxygen vacancy is lower in energy by 0.07 eV than that with equatorial oxygen vacancy (see Figure S5).

## 3. Conclusion

To summarize, the above experimental data demonstrate domain reproducibility around the paraelectric-ferroelectric and ferroelectric-ferroelectric phase transformations in the seminal ferroelectric barium titanate. The atomic-scale observations and modeling indicate that the origin of domain repeatability stems from immobile oxygen vacancies. In the current work, the memory effect was observed when the samples were heated up to 180 C and cooled back.  The memory effect was sustainable for time scales as long as even 10 hours. These values reflect temperatures much higher than $T_C$ and durations longer than the typical relaxation time in these materials [9,42–45]. Thus, while barium titanate is considered a ferroelectric material that undergoes an order-disorder phase transition, it is reasonable to assume that similar effects exist in materials that experience a displacive phase transformation. This assumption is backed up by the fact that a memory effect of the domain distribution is widely reported [47-48] for materials of these two families, in which domains relax to their original distribution after electric-field excitations, in parallel to the thermal excitation in the current work. Another point to clarify is the effect of surface chemistry on the memory effect. In principle, oxygen atoms and oxygen vacancies are expected to be more dynamic near the surface [54]. However, there are several good reasons to believe that the memory effect reported here is general. First, while AFM and PFM are more sensitive to surface, elastic domains should have a trace in topography of at least one-unit cell, which is readily detectable with AFM. However, Figure 1 shows no signal to such domains above $T_C$. Second, the DFT calculations predict stable oxygen vacancies in a system similar to the experimental one without making any assumptions on the surface chemistry. Moreover, while TEM samples are of 50-100 nm in thickness, the imaging is done by averaging the signal throughout the entire sample. In addition, Figure S6 shows optical micrographs that were taken by an optical objective that is placed in the AFM below and above $T_C$, supporting the bulk origin of the effect. It is also worth noting that the domain distribution is not perfectly repeatable as some changes in domain-wall position, mainly related to the domain width (and less so for the orientation) does have some flexibility. These results are in agreement with the TEM observations that show that not all oxygen vacancies have remained pinned after the phase transition. In addition, one should bear in mind that the common orientation of neighboring domains with a bundle domain [41,52] or a superdomain [53] contributes to the reduced randomness in domain distribution. That is, high correlation between domain re-distribution is possible even with a small number of oxygen vacancies, because even if the vacancies pin a small number of domain walls, the neighboring domain walls are affected by this pinning.

The immobile oxygen vacancies reported in this work experimentally and theoretically can explain the origin for the longstanding and partially understood existence of a polar phase that persists into the paraelectric phase, *e.g.*, as in the case of persistent pyroelectricity in the cubic phase of $BaTiO_3$ that was first reported by Chynoweth [24,46–48]. Lastly, the repeating domain redistribution presented here can explain the macroscopic shape-memory effect that was observed in several ferroelectrics [17,18,25,49,50], and even in non-polar ferroelastic materials [51], rendering ferroelectrics and other metal-oxides for multiscale shape-memory applications.

## Methods

**Density Functional Theory (DFT):** The DFT modeling was done using a periodically repeated 2×2×2 supercell which contains 8 Ba, 8 Ti and 23 O atoms, corresponding to a deficient $BaTiO_{3-x}$ ($x$=0.125). The existence of a stable oxygen vacancy at crystallographic phases that are energetically favorable above the absolute-zero temperature was discussed in detailed elsewhere [31]. The minimum energy paths of oxygen-vacancy diffusions in tetragonal and cubic $BaTiO_3$ were determined using the nudged elastic band (NEB) technique implemented in the USPEX code with lattice constants fixed to their corresponding bulk values. The plane-wave cutoff is set to 400 eV. A 2×2×2 Monkhorst-Pack $k$-point grid was used for structural optimizations and NEB calculations. The root-mean-square forces on images smaller than 0.03 eV/Å was the halting criteria condition for NEB calculations. The variable elastic constant scheme was used, and the spring constant between the neighboring images was set in the range of 3.0 to 6.0 eV/Å$^2$. First-principles density functional theory calculations were performed using Vienna Ab initio simulation package (VASP) with generalized gradient approximation of the Perdew-Burke-ErnZerhof for solids (PBEsol) type [35,36]. Optimized lattice constants of $BaTiO_3$ for the tetragonal phase $a$=$b$=3.957 Å, and $c$=4.033 Å and the cubic phase $a$=$b$=$c$=3.970 Å were considered.

**Transmission Electron Microscopy (TEM):** Single-crystal $BaTiO_3$ nanoparticles of 50 nm size purchased from US Research Nanomaterials, Inc. (99.9% pure) were used [7,30–32]. The particles were suspended in ethanol and sprayed on the DENS silicon chip with SiN film. The STEM-iDPC experiments were carried out in an aberration-corrected Titan Themis 80–300 operated at 200 kV, with a 30 - 50 pA dose [7, 31].

**Atomic Force Microscopy (AFM):** Single crystals of 5×5×1 mm$^3$ $BaTiO_3$ were used. Samples (three) were obtained from MaTecK [9] and were cleaned with acetone and ethanol. AFM and PFM measurements were carried out with an MFP-3D AFM, Asylum Research Ltd. with a diamond-coated tip (Adama, 1.56 N/m). PFM measurements were carried out while driving 15 V ac voltage at 197 kHz for the vertical signal and 6 V ac voltage at 603 kHz for the lateral signal. PFM and AFM images were taken at 512×512 and 256×256 pixel-resolution scans. For all images, both the backward and forward scans were recorded, while here, only the forward-direction images are presented for the sake of consistency. WSxM [9] was used to analyse the data.

## Acknowledgments

The Technion team acknowledges support from the Zuckerman STEM Leadership Program, the Technion Russel Barry Nanoscience Institute and Pazy Research Foundation Grant No. 149-2020 that supported also S.C. We also thank Dr. Yaron Kauffman and Mr. Michael Kalina for technical support. L.M. and S.L. acknowledge the support from Westlake Education Foundation. The computational resource is provided by Westlake HPC Center.

# Supplementary Information for "***Oxygen-vacancy Mediated Deterministic Domain Distribution at the Onset of Ferroelectricity***"

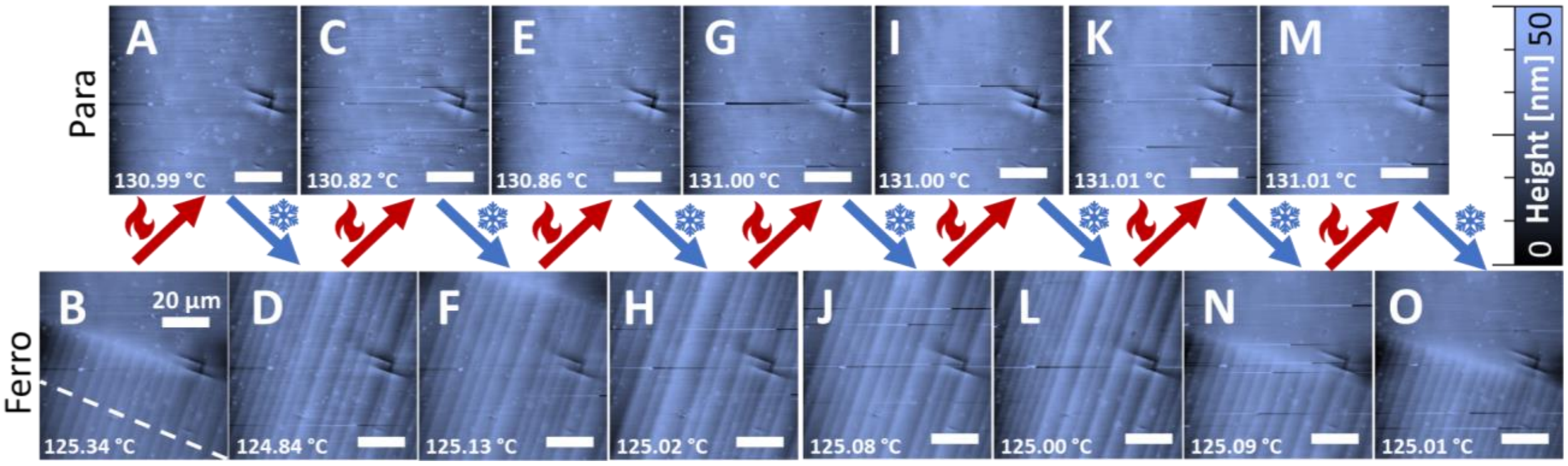

**Figure S1| Domain-memory during periodic paraelectric-ferroelectric phase-transition cycles.** (**A-N**) Domain distribution (AFM micrographs) above (no domains in A, C, E, G, I, K, M) and below (stripes in B, D, F, H, J, L, N and O) $T_C$ during eight phase transformations cycles showing a similar pattern. Temperature-variation path is designated by arrows, whereas the scan temperature is given for each micrograph. Note that the domain distribution in some of these micrographs was imaged when the material only partially transferred from cubic to tetragonal. Thus, the border between the tetragonal domain structure (stripes in the bottom part of the images) and the cubic phase with no domains varies between the images. Paraelectric-ferroelectric transitions represented by images A-D, E-H, G-J are used in Figure 3. Highest correlation (Figure 4) to all sister distributions was found for (B). Line-profile shown in (B) shows the cross section that was used to calculate the correlation coefficient with the 10,000 simulated signals (Figure 4B).

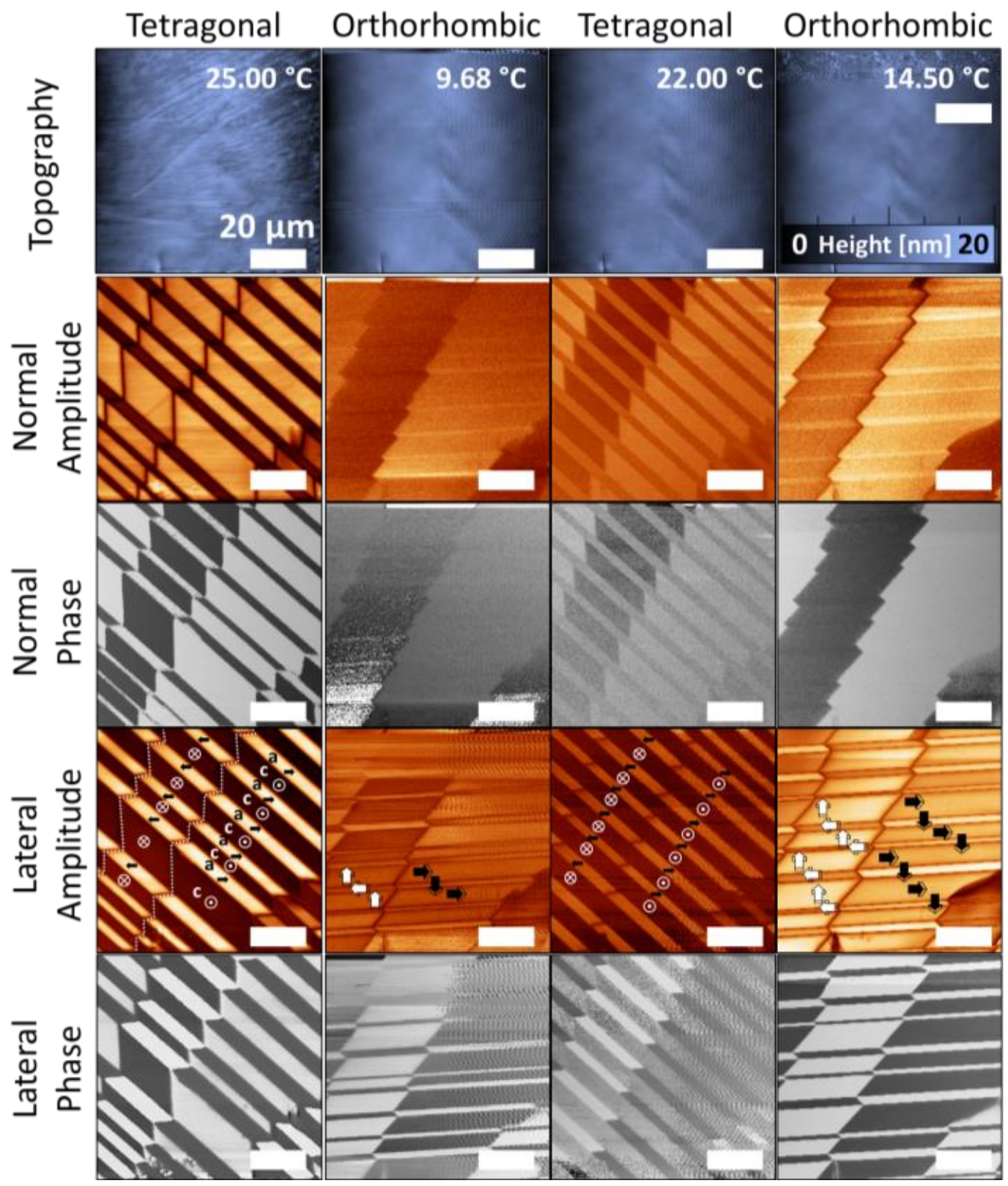

**Figure S2| Domain repeatability during ferroelectric-ferroelectric phase-transition cycles.** Domain repeatability during two ferroelectric-ferroelectric (left-to-right: tetragonal-orthorhombic-tetragonal-orthorhombic) phase-transition cycles revealed with (top-to-bottom:) simultaneous images of topography, vertical amplitude, vertical phase, lateral amplitude and lateral phase. The scan temperatures are given in the topography micrographs. The lateral-amplitude images are displayed in Figure 5.

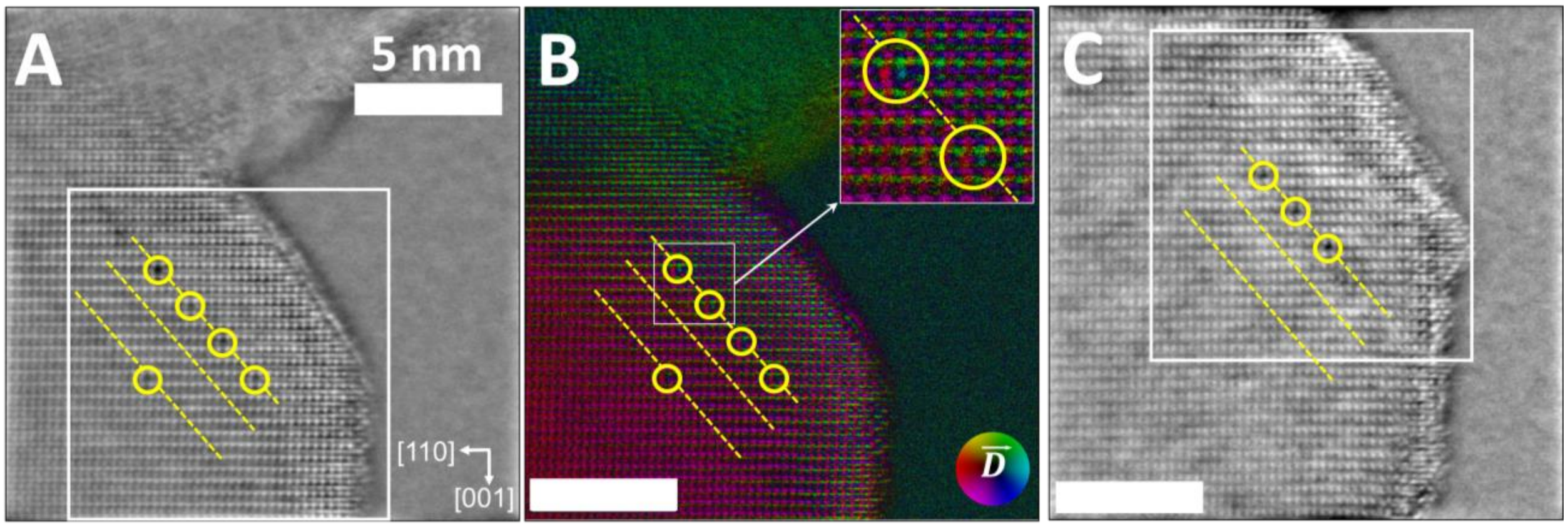

**Figure S3| Oxygen-vacancy-pinning during paraelectric-ferroelectric transition.** (**A**) Large-scale iDPC and (**B**) DPC images of the area that includes Figure 2A (designated in a square), showing the presence of domain walls (marked in dashed lines), which comprise oxygen vacancies (highlighted with circles). (**C**) A large-scale iDPC micrograph of the same area as in Figure 2B.

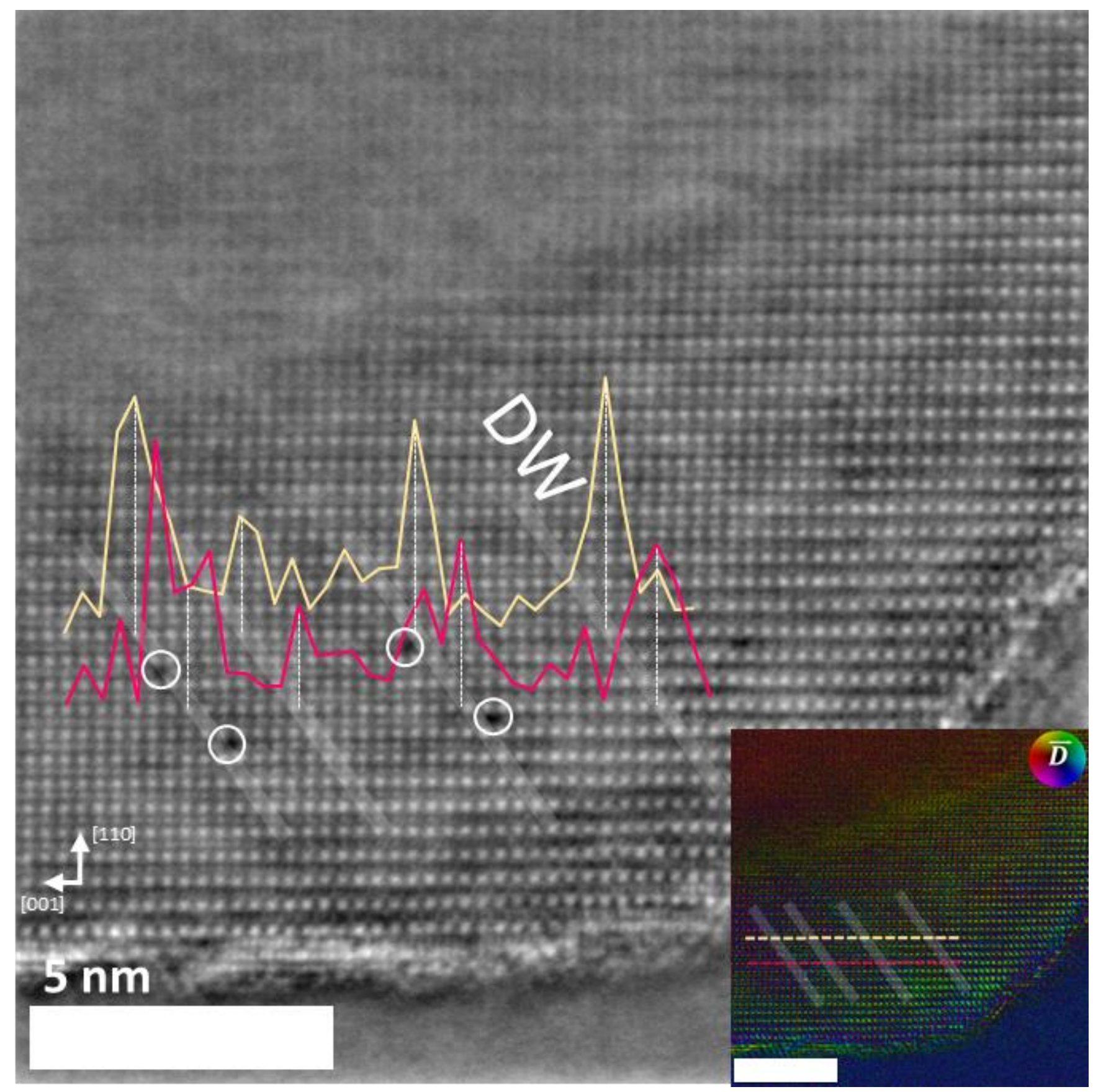

**Figure S4| Domain wall and oxygen vacancy positions.** iDPC image showing the location of oxygen vacancies (marked with white circles). Yellow and magenta graphs show the difference in hue values at the simultaneously imaged DPC image (insert) along the corresponding cross-sections (designated in the insert). The peaks in these graphs represent domain walls (example is denoted with DW), revealing that the oxygen vacancies are placed at the domain walls. White diagonal lines show the domain-wall position. Color wheel in the insert represents the direction and the intensity of the electric field displacement vector.

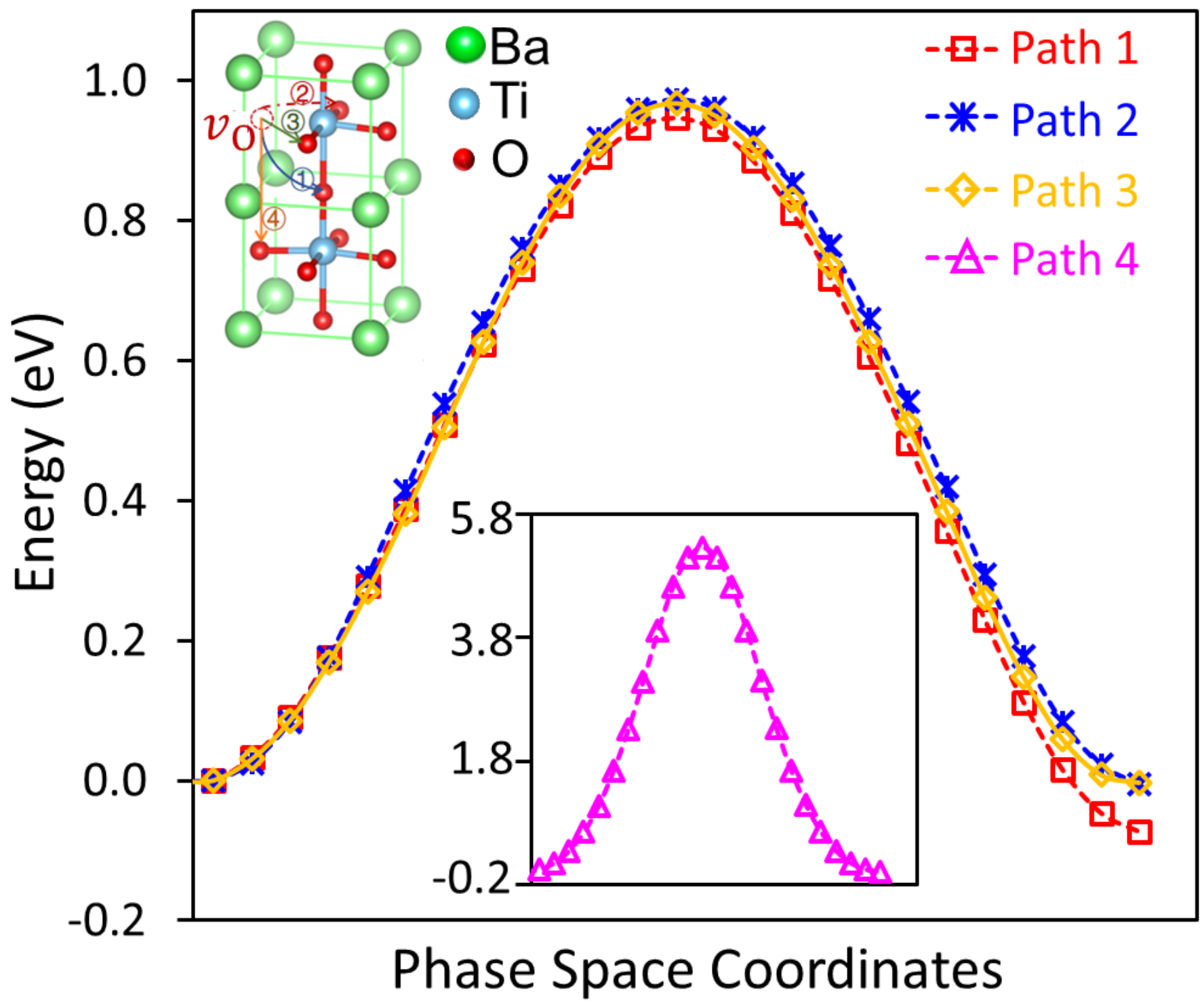

**Figure S5| Energy profiles for oxygen-vacancy diffusion in tetragonal $BaTiO_3$.** Profiles for various paths for oxygen-vacancy diffusion. Path 1 (lowest barrier) is the path shown in Figure 1.

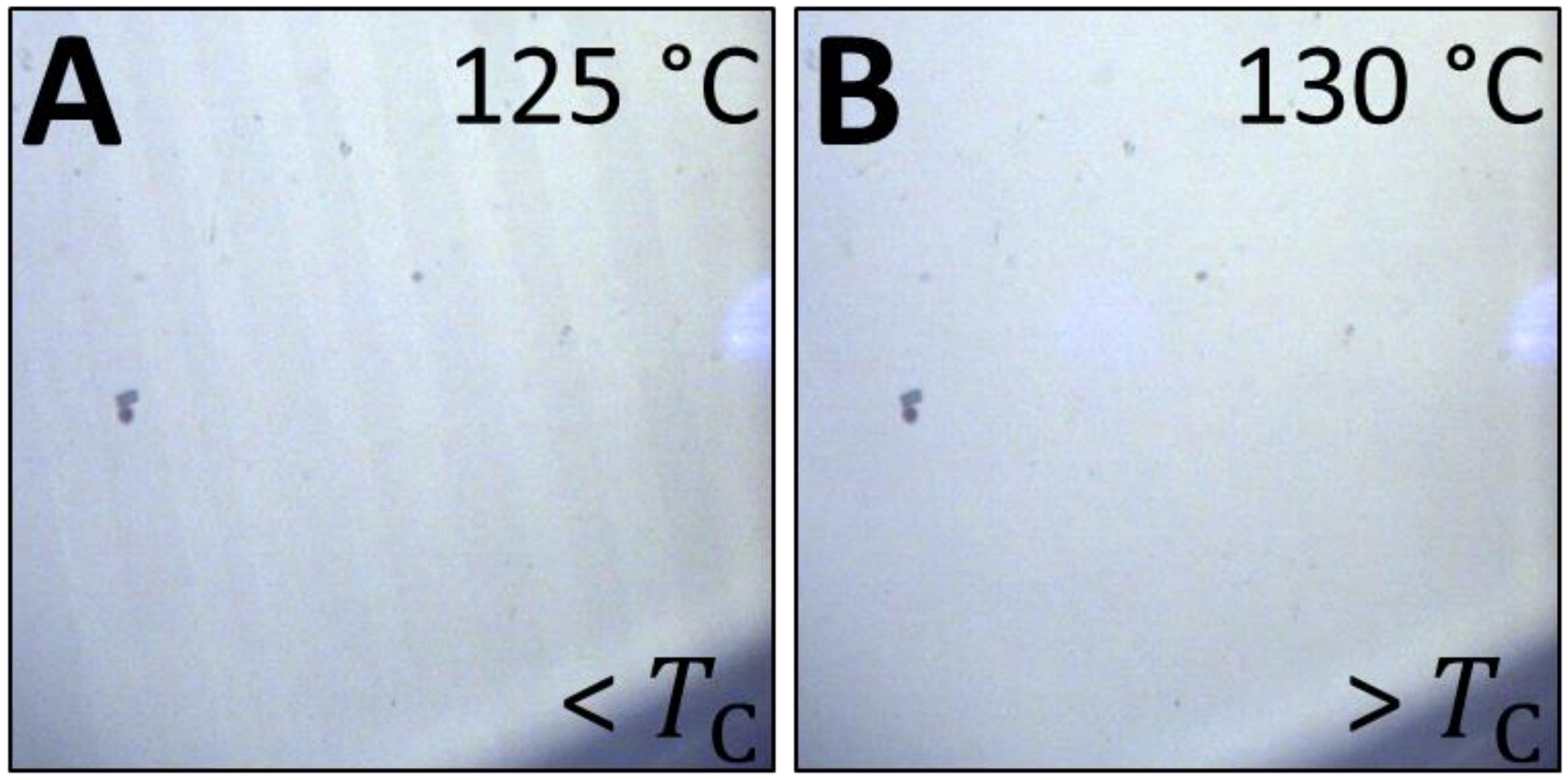

**Figure S6| Optical micrographs during ferroelectric-paraelectric phase transition cycle.** (**A**) Domains at the ferroelectric state and (**B**) the absence of domains in the same area and under the same imaging conditions, but above $T_C$. The image temperature is written as insets. Image size is ~40 x 40 μm$^2$.